\documentclass[twocolumn,showpacs,preprintnumbers,amsmath,amssymb]{revtex4}
\usepackage[dvips]{graphicx}
\usepackage{epsfig}
\usepackage{soul}

\begin{document}

\title{Freezing dynamics of entanglement and nonlocality for qutrit-qutrit ($3 \otimes 3$) quantum systems}
\author{Mazhar Ali}
\affiliation{Department of Electrical Engineering, Faculty of Engineering, Islamic University Madinah, 107 Madinah, Saudi Arabia}

\begin{abstract}
We examine the possibilities of non-trivial phenomena of time-invariant entanglement and freezing dynamics of entanglement for qutrit-qutrit 
quantum systems. We find no evidence for time-invariant entanglement, however, we do observe that quantum states freeze their entanglement after 
decaying for some time. It is interesting that quantum states are changing whereas their entanglement remains constant. We find that the combined action 
of decoherence free subspaces and subspaces where quantum states decay, facilitate this phenomenon. This study is an extension of similar phenomena 
observed for qubit-qubit systems, qubit-qutrit, and multipartite quantum systems. We examine nonlocality of a specific family of states and find the certain 
instances where the states still remain entangled, however they can either loose their nonlocality at a finite time or remain nonlocal for all times.
\end{abstract}

\pacs{03.65.Yz, 03.65.Ud, 03.67.Mn}

\maketitle

\section{Introduction}
\label{Sec:intro}

Entanglement and nonlocality are two features of quantum mechanics which have attracted lot of interest and considerable 
efforts have been devoted to develop a theory of these phenomenona \cite{Horodecki-RMP-2009, gtreview}. 
Due to growing efforts for experimental realizations of devices utilizing these features, it is essential to study the effects of noisy environments 
on entanglement and nonlocality. Such studies are an active area of research \cite{Aolita-review} and several authors have studied decoherence effects 
on quantum correlations for both bipartite and multipartite systems
\cite{Yu-work,lifetime,Aolita-PRL100-2008,bipartitedec,Band-PRA72-2005,lowerbounds,Lastra-PRA75-2007, Guehne-PRA78-2008,Lopez-PRL101-2008, 
Ali-work, Weinstein-PRA85-2012,Ali-JPB-2014, Ali-2015, Ali-2016, Ali-2017}. 

The specific noise dominant in experiments on trapped atoms is caused by intensity fluctuations of electromagnetic fields which leads 
to collective dephasing process. The dynamics of entanglement under collective dephasing has been studied for both bipartite and multipartite quantum 
systems \cite{Yu-CD-2002, AJ-JMO-2007,Li-EPJD-2007, Karpat-PLA375-2011, Liu-arXiv, Carnio-PRL-2015, Carnio-NJP-2016, Song-PRA80-2009, Ali-PRA81-2010}.  
Some of these previous studies \cite{Ali-2017, Karpat-PLA375-2011, Liu-arXiv, Carnio-PRL-2015, Carnio-NJP-2016}, revealed 
two interesting features of the dynamical process, which are so called {\it freezing} dynamics of entanglement \cite{Ali-2017} and 
{\it time-invariant} entanglement \cite{Karpat-PLA375-2011, Liu-arXiv, Carnio-PRL-2015, Carnio-NJP-2016}. It was shown that a specific two qubits 
state may first decay upto some numerical value before suddenly stop decaying and maintain this stationary entanglement \cite{Carnio-PRL-2015}. 
Such behavior was also observed for various genuine multipartite specific states of three and four qubits, including random states \cite{Ali-2017}.  
Other interesting dynamical feature under collective dephasing is the possibility of {\it time-invariant} entanglement.  
Time-invariant entanglement does not necessarily mean that the quantum states live in decoherence free subspaces (DFS). In fact the quantum states 
may change at every instance whereas their entanglement remain constant throughout the dynamical process. This feature was first observed for 
qubit-qutrit systems \cite{Karpat-PLA375-2011} and more recently for qubit-qubit systems \cite{Liu-arXiv}. 
We have investigated time-invariant phenomenon for genuine entanglement of three and four qubits. We have explicitly observed this phenomenon for a 
specific family of quantum states of 
four qubits \cite{Ali-2017}. For qutrit-qutrit ($3 \otimes 3$) systems, some features of entanglement dynamics under collective dephasing are 
known, in particular the phenomenon of distillability sudden death \cite{Song-PRA80-2009, Ali-PRA81-2010}, however, so far to our knowledge, 
the possibility of time-invariant entanglement and freezing dynamics of entanglement has not been studied so far. In this work, we investigate these 
two features for this dimension of Hilbert space for a specific family of states and also for some random states. 

Another aspect of quantum correlations is quantum nonlocality, which refers to the phenomenon that the predictions made using quantum 
mechanics cannot be simulated by a local hidden variable model. The existance of nonlocal correlations can be detected via violation of some types of Bell 
inequalities \cite{Bell-Phys-1964}. It is well known that pure entangled states violate a Bell inequality, whereas mixed entangled states may not do 
so \cite{Gisin-Werner-1991}. It is also known that entangled states do exhibit some kind of hidden nonlocality \cite{Liang-PRA86-2011}. The well known 
Clauser-Horne-Shimony-Holt (CHSH) inequality \cite{CHSH-1969} for two qubits has been studied under decoherence both in 
theory \cite{Mazzola-PRA81-2010}, and experiment \cite{Xu-PRL-2010}. 
Several investigations of nonlocality of multipartite quantum states under decoherence have been carried out \cite{NLD}.
The extension of CHSH inequality for multipartite quantum systems has received considerable attention in theory 
\cite{Mermin-PRL-1990, Ardehali-PRA-1992, Bancal-PRL-2011, Svetlichny-PRD-1987, Collins-PRL88-2002, Chen-PRA64-2001} and 
in experiments \cite{Pan-expMBI, Bastian-PRL104-2010}. We have recently studied the effect of collective dephasing on genuine nonlocality of 
quantum states exhibiting time-invariant and freezing entanglement dynamics \cite{Ali-2017}. The problem of quantum nonlocality for high dimensional systems 
has been studied \cite{Kaszlikowski-PRL-2000, Kaszlikowski-PRA65-2002, Chen-PRA74-2006}. One particular inequality is called 
Collins-Gisin-Linden-Massar-Popescu (CGLMP) inequality \cite{CGLMP-PRL-2002}. In this work, we also study the effect of collective dephasing on 
nonlocality of qutrit-qutrit systems using CGLMP inequality.

This paper is organized as follows. In section \ref{Sec:Model}, we briefly discuss our model of interest. In section \ref{Sec: EnN}, we review the idea of 
maximally entangled states for qutrit-qutrit systems and describe the method to compute a specific measure of entanglement for an arbitrary initial 
quantum state. We also review nonlocality and its computation in the same section. In section \ref{Sec:res}, we provide our main results. Finally, 
we conclude our work in section \ref{Sec:conc}.

\section{Collective dephasing for qutrit-qutrit systems} 
\label{Sec:Model}

Our physical model consists of two qutrits (two three-level atoms for example) $A$ and $B$ that are coupled to a noisy environment,  
collectively. Our qutrits are sufficiently far apart and they do not interact with each other, so that we can treat them as independent. The 
collective dephasing refers to coupling of qutrits to the same noisy environment $B(t)$.
The Hamiltonian of the quantum system (with $\hbar = 1$) can be written as \cite{AJ-JMO-2007, Ali-PRA81-2010} 
\begin{eqnarray}
H(t) = - \frac{\mu}{2} \, \big[ \, B(t) (\sigma_z^A + \sigma_z^B) \,  \big] \, , \label{Eq:Ham} 
\end{eqnarray}
where $\mu$ is gyromagnetic ratio and $\sigma_z $ denotes the dephasing operator for qutrits $A$ and $B$. The stochastic magnetic fields refer 
to statistically independent classical Markov processes satisfying the conditions
\begin{eqnarray} 
\langle B(t) \, B(t')\rangle &=& \frac{\Gamma}{\mu^2} \, \delta(t-t') \,, \nonumber \\ 
\langle B(t)\rangle &=& 0 \, ,
\end{eqnarray}
with $\langle \cdots \rangle$ as ensemble time average and $\Gamma$ denote the phase-damping rate for collective decoherence.

Let $|2\rangle$, $|1\rangle$, and $|0\rangle$ be the first excited state, second excited, and ground state of the qutrit, respectively. 
We choose the computational basis $ \{ \, |0,0\rangle$, $|0,1\rangle$, 
$|0,2\rangle \, |1,0\rangle$, $|1,1\rangle$, $|1,2\rangle$, $|2,0\rangle$, $|2,1\rangle$, $|2,2\rangle$,$\}$, where we have dropped the 
subscripts $A$ and $B$ with the understanding that first basis represents qutrit $A$ and second qutrit $B$. Also the notation 
$|0 \rangle \otimes |0\rangle = |0 \, 0 \rangle$ has been adopted for simplicity. 
The time-dependent density matrix for two-qutrits system is obtained by taking ensemble average over the noisy field,
i.\,e., $\rho(t) = \langle\rho_{st}(t)\rangle$, where $\rho_{st}(t) = U(t) \rho(0) U^\dagger(t)$ and
$U(t) = \exp[-\mathrm{i} \int_0^t \, dt' \, H(t')]$. The dynamics of density matrix can be given by operator sum 
representation \cite{AJ-JMO-2007} as $\rho(t) = \sum_j^n \, K_j^\dagger(t) \rho(0) K_j(t)$, where $K_j$ are 
Kraus operators that preserve the positivity and unit trace conditions, i.\,e., $\sum_j^n \, K_j^\dagger K_j = \mathbb{I}$. 
The most general solution of $\rho(t)$ under the assumption that the system is 
not initially correlated with environment is given as
\begin{eqnarray}
\rho(t) = \, \sum_{k=1}^3 \, (D_k^{AB \, \dagger} \, \rho(0) \, D_k^{AB})\, , \label{Eq:GS}
\end{eqnarray}
where the terms describing interaction with collective magnetic field $B(t)$ involve the operators 
$D_1^{AB} = \mathrm{diag} (\gamma(t), 1,1,1,\gamma(t) \,,1,1,1,\gamma(t))$, 
$D_2^{AB} = \mathrm{diag} (\omega_1(t), 0,0,0,\omega_2(t) \,,0,0,0,\omega_2(t))$, and
$D_3^{AB} = \mathrm{diag} (0,0,0,0,\omega_3(t),0,0,0,\omega_3(t))$. 
The time dependent parameters are defined as, $\gamma = \mathrm{e}^{-\Gamma t/2}$, $\omega_1(t) = \sqrt{1-\gamma^2(t)}$, 
$\omega_2(t) = - \gamma^2(t) \, \sqrt{1-\gamma^2(t)}$, and $\omega_3(t) = (1-\gamma^2(t)) \, \sqrt{1+\gamma^2(t)}$.

The matrix form of Eq.~(\ref{Eq:GS}) for an arbitrary initial state is given as
\begin{widetext}
\begin{eqnarray}
\rho(t) = \left(
\begin{array}{lllllllll}
\rho_{11} & \gamma \rho_{12} & \gamma \rho_{13} & \gamma \rho_{14} & \gamma^4 \rho_{15} & \gamma \rho_{16} & \gamma \rho_{17} & \gamma \rho_{18} & \gamma ^4 \rho_{19} \\
\gamma \rho_{21} & \rho_{22} & \rho_{23} & \rho_{24} & \gamma \rho_{25} & \rho_{26} & \rho_{27} & \rho_{28} & \gamma \rho_{29} \\
\gamma \rho_{31} & \rho_{32} & \rho_{33} & \rho_{34} & \gamma \rho_{35} & \rho_{36} & \rho_{37} & \rho_{38} & \gamma \rho_{39} \\
\gamma \rho_{41} & \rho_{42} & \rho_{43} & \rho_{44} & \gamma \rho_{45} & \rho_{46} & \rho_{47} & \rho_{48} & \gamma \rho_{49} \\
\gamma^4 \rho_{51} & \gamma \rho_{52} & \gamma \rho_{53} & \gamma \rho_{54} & \rho_{55} & \gamma \rho_{56} & \gamma \rho_{57} & \gamma \rho_{58} & \rho_{59} \\
\gamma \rho_{61} & \rho_{62} & \rho_{63} & \rho_{64} & \gamma \rho_{65} & \rho_{66} & \rho_{67} & \rho_{68} & \gamma \rho_{69} \\
\gamma \rho_{71} & \rho_{72} & \rho_{73} & \rho_{74} & \gamma \rho_{75} & \rho_{76} & \rho_{77} & \rho_{78} & \gamma \rho_{79} \\
\gamma \rho_{81} & \rho_{82} & \rho_{83} & \rho_{84} & \gamma \rho_{85} & \rho_{86} & \rho_{87} & \rho_{88} & \gamma \rho_{89} \\
\gamma ^4 \rho_{91} & \gamma \rho_{92} & \gamma \rho_{93} & \gamma \rho_{94} & \rho_{95} & \gamma \rho_{96} & \gamma \rho_{97} & \gamma \rho_{98} & \rho_{99}
\end{array}
\right) \,.\label{Eq:MF}
\end{eqnarray}
\end{widetext}
We note that decoherence free subspaces (DFS) \cite{Yu-CD-2002} do appear in this system as a common characteristic of collective dephasing. 
Another interesting property of the dynamics is the fact that all initially zero matrix elements remain zero.

\section{Entanglement and nonlocality for $3 \otimes 3$ quantum systems}
\label{Sec: EnN}

In this section, we briefly review the key ideas and certain work related with entanglement and nonlocality for qutrit-qutrit systems. In 
subsection \ref{SubSec:RIQS}, we discuss the maximally entangled states and a computable measures of entanglement. In 
subsection \ref{SubSec:nonlocality}, we review the nonlocality and methods to quantify it for quantum states.  

\subsection{Maximally entangled states of qutrit-qutrit systems}
\label{SubSec:RIQS}

The analog of Bell-diagonal states of two qubits for qutrit-qutrit systems is simplex \cite{Baumgartner-PRA-2006}, which lives in nine dimensional real 
linear space. Let us consider the maximally entangled pure state, given as  
\begin{eqnarray}
|\Psi_{0,0} \rangle = \frac{1}{\sqrt{3}} \sum_{k = 0}^2 |k\rangle \otimes |k \rangle) \,.
\label{Eq:ME1}
\end{eqnarray}
We can construct the basis of $\mathbb{C}^3 \otimes \mathbb{C}^3$ consisting of maximally entangled pure states, like Bell states as follows. 
Let $\mathbb{M}$ is set of indices $(m,n)$, where $m,n \in \mathbb{Z}_3$ with addition and multiplication of indices as modulo 3. 
For each pair $\lambda = (m,n) \in \mathbb{M}$, we can define a unitary operator 
\begin{equation}
 W_\lambda = W_{(m,n)}  = \sum_{k = 0}^2 e^{\frac{2 \, \pi \, i}{3} kn} \, |k\rangle \langle k + m|\,.
\end{equation}
Then to each point $\lambda \in \mathbb{M}$, we associate the vector $|\Psi_\lambda \rangle \in \mathbb{C}^3 \otimes \mathbb{C}^3$, as 
\begin{equation}
| \Psi_\lambda \rangle= (W_\lambda \otimes \mathbb{I}) \, |\Psi_{0,0}\rangle \, .
\label{Eq:MEW}
\end{equation}
So we obtain the nine maximally entangled vectors, which form a basis of two qutrits vector space. We note that for collective dephasing model, 
six of these states reside in DFS and the rest of the 3 states reside in a space which is decoupled from DFS. 
While the geometry of Bell-diagonal states can be considered
as tetrahedron with Bell states sitting at four corners, the corresponding geometry for qutrits is not that intuitive \cite{Baumgartner-PRA-2006}, however 
both cases contain the maximally mixed state ``$ \mathbb{I}/N^2$'' at the center of tetrahedrons. 

The problem of detection and quantification of entanglement for qutrit-qutrit systems is not an easy one. It is well know that for a given bipartite 
quantum state, if the matrix with a partial transpose taken with respect to either of the subsystem, has at least one negative eigenvalue, then the 
quantum state is entangled and called NPT state \cite{Peres-PRL-1996}. There exist some qutrit-qutrit quantum states which have positive partial 
transpose (PPT), nevertheless, they are entangled \cite{Horodecki-RMP-2009, Clarisse-PhD}. These PPT-entangled states or so called bound entangled 
states (BES) pose the 
actual difficulty in order to characteristic and quantify entangled states. Although there are few criteria, like realignment criterion also called 
cross-norm criterion \cite{Chen-QIP-2003} to detect some of bound entangled states, nevertheless, in general the problem of detection of 
entanglement for this dimension of Hilbert space is not solved and it is an open issue. 

Negativity \cite{Vidal-PRA65-2002} is a useful measure to quantify entanglement of bipartite NPT states, however, strictly speaking, this measure 
do not captures the bound entangled states, and for a given initial state, if negativity is zero, then it is not known in general whether the states 
are entangled or not except for isotropic states. In addition, as the dimension of 
Hilbert space for bipartite systems is larger than $4$ (qubit-qubit system), it is not always easy to find the analytical expressions for eigenvalues for 
an initial state. As negativity is quantified by adding negative eigenvalues, so exact expression for it is also intractable in many cases. Even 
the numerical computations could be done but the procedure is also not straightforward. The difficulty in computing the negativity numerically is 
now removed by recent studies in quantification of multipartite entanglement \cite{Bastian-PRL106-2011}. Although the main efforts of 
authors \cite{Bastian-PRL106-2011} were to quantify genuine entanglement for multipartite quantum systems, nevertheless, the genuine negativity 
simply gives the usual measure of negativity for bipartite quantum systems. We have used this measure in our study. The further details on computing 
this measure can be found in original work \cite{Bastian-PRL106-2011}.   

\subsection{Quantum nonlocality for qutrits}
\label{SubSec:nonlocality}

For a brief description of bipartite nonlocality, consider that each party can perform a measurement $X_j$ with result $a_j (b_j)$ for $j = A(B)$. The 
joint probability distribution $P(a_A b_B |X_A X_B) $ may exhibit different notions of nonlocality. It may be that it cannot be written 
in local form as
\begin{equation}
 P(a_A b_B |X_A X_B) = \int d\lambda P_A(a_A|X_A \lambda) \, P_B(b_B|X_B \lambda)\, ,
\label{Eq:PLV}
 \end{equation}
where $\lambda$ is a shared local variable. Such nonlocality can be tested by standard Bell inequalities. One such inequality for two parties, two settings, 
and three outcomes is called CGLMP inequality \cite{CGLMP-PRL-2002}. The experimental violation of this inequality has been 
observed \cite{Vaziri-PRL-2002} as well. The inequality is given as
\begin{eqnarray}
P(a = b) + P(b = a' + 1) + P(a' = b') \nonumber \\ + P(b' = a) - P(a = b -1) - P(b = a') \nonumber \\ - P(a' = b' - 1) - P(b' = a-1) \leq 2 \,, 
\label{Eq:CGLMP}
\end{eqnarray}
where the outcomes are $0,1,2$ and sum inside probabilities are modulo 3. The Bell operator associated with this inequality can be 
written \cite{Alsina-PRA94-2016} as
\begin{eqnarray}
 \mathcal{B} = 2 - 3 (a^2 + b'^2) + \frac{3}{4} (ab + a^2 b - a' b -a'^2 b \nonumber \\
               - a b^2 + a' b^2 + ab' -a^2 b' + a'b' + a'^2 b' + ab'^2 \nonumber \\ 
               - a'b'^2) + \frac{9}{4} (a^2 b^2 - a'^2 b^2  + a^2 b'^2 + a'^2 b'^2) \, ,
\label{Eq:BOG}
\end{eqnarray}
where we have omitted the tensor product symbols. The optimal measurements can be expressed in terms of eight Gell-Mann matrices. In general the 
exact form of these measurements are initial state dependent. However for an initial entangled state of the form $|\Psi_{0,0}\rangle$, the Bell operator 
\cite{Acin-PRA65-2002} is given as 
\begin{eqnarray}
\mathcal{B} = \left(
\begin{array}{lllllllll}
0 & 0 & 0 & 0 & \frac{2}{\sqrt{3}} & 0 & 0 & 0 & 2 \\
0 & 0 & 0 & 0 & 0 & \frac{2}{\sqrt{3}} & 0 & 0 & 0 \\
0 & 0 & 0 & 0 & 0 & 0 & 0 & 0 & 0 \\
0 & 0 & 0 & 0 & 0 & 0 & 0 & \frac{2}{\sqrt{3}} & 0 \\
\frac{2}{\sqrt{3}} & 0 & 0 & 0 & 0 & 0 & 0 & 0 & \frac{2}{\sqrt{3}} \\
0 & \frac{2}{\sqrt{3}} & 0 & 0 & 0 & 0 & 0 & 0 & 0 \\
0 & 0 & 0 & 0 & 0 & 0 & 0 & 0 & 0 \\
0 & 0 & 0 & \frac{2}{\sqrt{3}} & 0 & 0 & 0 & 0 & 0 \\
2 & 0 & 0 & 0 & \frac{2}{\sqrt{3}} & 0 & 0 & 0 & 0 \\
\end{array}
\right) \,.\label{Eq:BOS}
\end{eqnarray}
It is well known that entangled qutrits violate local realism more stronger than qubits \cite{Kaszlikowski-PRL-2000}. For qubits, the maximum violation is 
achieved by Bell states and it is equal to $2\,\sqrt{2} \approx 2.8284$, whereas for qutrits the violation by maximally entangled state is equal to 
$\approx 2.8729$ \cite{Acin-PRA65-2002}. Surprisingly, it was found that the maximum violation for two qutrits is not achieved by maximally entangled 
state but by a non-maximally entangled state \cite{Acin-PRA65-2002} given as
\begin{equation}
|\psi_\mu \rangle = \frac{1}{\sqrt{2 + \mu^2}} \, (|00\rangle + \mu |11\rangle + |22\rangle)\,, 
\end{equation}
with $\mu = (\sqrt{11} - \sqrt{3})/2 \approx 0.7923$. It is not difficult to check that 
\begin{equation}
\langle \, \mathcal{B} \, \rangle_{|\psi_\mu \rangle\langle \psi_\mu|} = \frac{12 + 8 \sqrt{3} \mu}{6 + 3 \mu^2}\,,    
\end{equation}
which is equal to $ \langle \, \mathcal{B} \, \rangle \approx 2.9149$ for $\mu \approx 0.7923$. Similarly, for another entangled state \cite{Acin-PRL-2005}
\begin{equation}
|\psi_\nu \rangle = \nu (|00\rangle + |11\rangle) + \sqrt{1 - 2 \nu^2} |22\rangle)\,, 
\end{equation}
where $ 0 \leq \nu \leq 1/\sqrt{2}$, the expectation value of Bell operator is given as
\begin{equation}
\langle \, \mathcal{B} \, \rangle_{ |\psi_\nu \rangle\langle \psi_\nu|} = 
\frac{4 \, \nu \, \big(\sqrt{3} \, \nu + (3 + \sqrt{3}) \sqrt{1 - 2 \, \nu^2} \, \big)}{3}\,,    
\end{equation}
which achieve its maximum violation for $\nu \approx 0.617$ \cite{Acin-PRL-2005}.

\section{Dynamics of entanglement and nonlocality}
\label{Sec:res}

All previous studies \cite{Karpat-PLA375-2011, Liu-arXiv, Carnio-PRL-2015, Carnio-NJP-2016, Ali-2017} which reported the possibility 
of {\it time-invariant} entanglement and {\it freezing} dynamics of entanglement have one fact common in their findings. The quantum states 
exhibiting these features are always mixtures of two entangled states and one of this entangled state reside in DFS such that when the entangled state 
living in subspace which is not decoherence free, decays, then somehow the entanglement present in DFS shields the combined states to preserve their 
entanglement. So, it is interesting to note that although quantum states (their eigenvalues as well) are changing all the time, nevertheless their 
entanglement remains stationary. In current system of two qutrits, we must also take a mixture of states from each subspace to check the possibility of 
time-invariant or freezing entanglement.  
To this aim, we define our initial states by mixing two entangled states from two decoupled subspaces. First let us consider the states, 
\begin{equation}
\rho_\alpha = \alpha |\Psi_{0,0}\rangle \langle \Psi_{0,0} | + \frac{1 - \alpha}{9}\, \mathbb{I}_9 \, ,    
\end{equation}
where $ 0 \leq \alpha \leq 1$. These states are called isotropic states and they have the property that their PPT region is separable \cite{Clarisse-PhD}. 
The states are NPT for $1/4 < \alpha \leq 1$, and hence entangled. These states exhibit nonlocality for 
$\alpha > 9/(2 (3 + 2 \sqrt{3})) \approx 0.6962$ \cite{Acin-PRA65-2002}. We now take another maximally entangled state residing in DFS, given as
\begin{equation}
|\Psi_{0,2}\rangle = \frac{1}{\sqrt{3}} \, \big( |02\rangle + |10\rangle + |21\rangle \, \big) \, .    
\end{equation}
This state is obtained using the relation (\ref{Eq:MEW}). We can now define a family of states, which are mixture of isotropic states and 
$|\Psi_{0,2}\rangle$, given as 
\begin{equation}
\rho_{\alpha,\beta} = \beta \, |\Psi_{0,2}\rangle \langle \Psi_{0,2} | + (1 - \beta) \, \rho_\alpha \, ,    
\end{equation}
where $ 0 \leq \beta \leq 1$. Although, the entanglement properties for this family of states may not be very clear, however, the spectrum of states do 
shed some light on their being NPT or PPT. The partial transpose w.r.t. subsystem $B$ gives $9$ eigenvalues and $6$ of them are definitely 
positive for the given range of parameters $\alpha$ and $\beta$. The $3$ possible negative eigenvalues are all same and given as
\begin{equation}
\frac{1}{9} \, \big[ (1-\alpha) (1-\beta) - 3 \sqrt{ \alpha^2 (1-\beta)^2 - \alpha \beta (1-\beta) + \beta^2 } \big] \, .    
\label{Eq:NEVs}
\end{equation}
It is not difficult to check that for $\beta > 1/4$, the states are NPT for $0 \leq \alpha \leq 1$. Alternatively, the states are again NPT if 
$\alpha > 0.277$ and $0 \leq \beta \leq 1$.

The time evolution of these states follows from Eq.~(\ref{Eq:MF}) and can be written as
\begin{equation}
\rho_{\alpha,\beta}(t) = \beta \, |\Psi_{0,2}\rangle \langle \Psi_{0,2} | + (1 - \beta) \, \rho_\alpha(t) \, .    
\end{equation}
Hence $\rho_\alpha(t)$ decays, whereas $|\Psi_{0,2}\rangle$ remain dynamically invariant as it lives in DFS. Now there is an additional 
parameter $\Gamma t$ involved in the density matrix and due to this parameter not all the eigenvalues of partially transposed matrix are tractable. 
However, an interesting observation is accessible, which is the fact that out of three possible negative eigenvalues (given in Eq.~(\ref{Eq:NEVs})), 
one eigenvalue remains same and independent of parameter $\Gamma t$. This simply implies that if we choose the parameters $\alpha > 0.277$ and 
$\beta > 1/4$, the time evolved density matrix $\rho_{\alpha, \beta}(t)$ remain NPT for all the time and hence remain entangled. 
The other two negative eigenvalues become unknown functions of parameters $(\alpha, \, \beta,\, \Gamma t)$ and do feed (vary) any reasonable 
measure of entanglement, like negativity for some time until their contribution becomes extremely small or zero.  

\begin{figure}[t!]
\scalebox{2.0}{\includegraphics[width=1.85in]{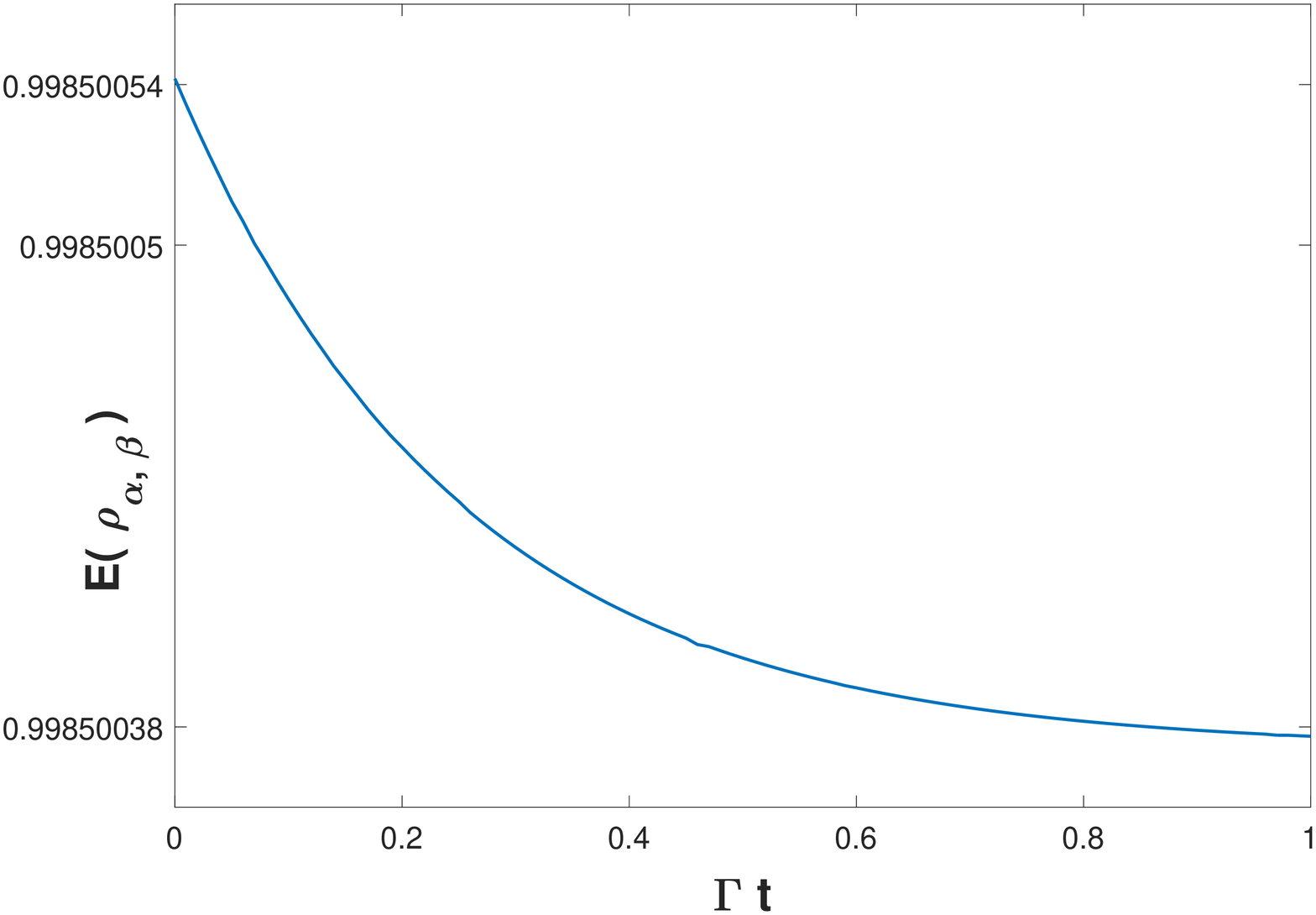}}
\caption{Entanglement monotone (negativity) for an initial state $\rho_{\alpha, \beta}(t)$ with parameters $\alpha = 0.999$ and $\beta = 0.999$. 
It can be seen that despite quite a very large fraction of states living in DFS, entanglement decays with an extremly small rate. See text for 
explanations.}
\label{Fig:1}
\end{figure}

Another observation in previous studies on time-invariant and freezing dynamics of entanglement is the fact that the fraction of states living in 
DFS must always be larger than other state. This makes sense as if the state living in decaying subspaces has larger probability in the mixture then 
due to decoherence of this fraction, entanglement is expected to decay faster. On the other hand if the state living in DFS has a larger fraction then its 
share of entanglement in combined state would also be larger and more stable. So only in these situations, one can expect either time-invariant 
entanglement or its freezing dynamics. This is exactly the expected dynamics which have found in Figure~(\ref{Fig:1}). Although, we have included 
a very tiny fraction of states living outside decoherence free subspaces, nevertheless the entanglement does not remain invariant and decays with a very 
small rate (in 7th place after decimal point.). This simple example indicates that actually there is no time-invariant entanglement for this 
dimension of Hilbert space. However, we do expect the freezing dynamics of entanglement as the curve tends towards value $0.9985$.   

\begin{figure}[t!]
\scalebox{2.0}{\includegraphics[width=1.85in]{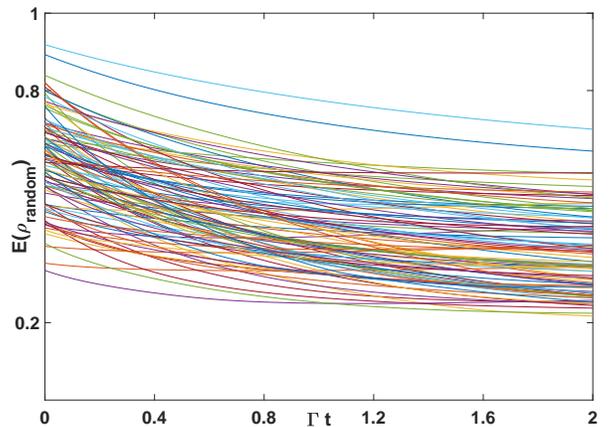}}
\caption{Entanglement monotone (negativity) is plotted against parameter $\Gamma t)$ for $100$ initial random pure states. It can be 
seen that all states remain NPT and more or less approach to a fixed (freezing) value of entanglement after some time.}
\label{Fig:RS2}
\end{figure}

In order to get a general trend of decay for an arbitrary initial states, we have generated $100$ random pure qutrit-qutrit states acoording to 
Haar measure \cite{Toth-CPC-2008}. We let these states interact with our reservoirs and we have studied their entanglement properties over a period of time. 
Figure~(\ref{Fig:RS2}) shows the results where it can seen that all of the initial NPT states remain NPT throughout the dynamical process. After initial 
entanglement decay upto a certain value, we observe the freezing behavior as expected. As all the off-diagonal elements living in DFS do not decay, so 
these elements in DFS retain their entanglement.

Finally, we study the CGLMP inequality for this state $\rho_{\alpha, \beta}(t)$. For this aim, we need an appropriate Bell operator which depends on 
the type of initial state we want to test. If we choose our parameters such that the off-diagonal term for DFS is larger than we need to apply unitary 
transformation to operator $\mathcal{B}$ (Eq.~(\ref{Eq:BOS})) before taking the expectation value of it, that is, 
\begin{equation}
\tilde{\mathcal{B}} = (W_{0,2} \otimes \mathbb{I}) \, \mathcal{B} \, (W_{0,2}^\dagger \otimes \mathbb{I}) \, ,
\label{Eq:BOT}
\end{equation}
where $W_{0,2}$ is the unitary operator to get state $|\Psi_{0,2}\rangle$ from state $|\Psi_{0,0}\rangle$. It is straightforward to calculate 
the expectation value of this operator, which is given as
\begin{equation}
\langle \, \tilde{\mathcal{B}} \, \rangle_{\rho_{\alpha, \beta}(t)} =  
\frac{4 \, \big(\sqrt{3} \, \alpha \, (1-\beta)\, e^{-2\, \Gamma t} + (3 + 2\, \sqrt{3}) \beta \, \big)}{9}\,.    
\end{equation}
It is clear that for $\beta = 1$, the state is trivially time-invariant (as it lives in DFS) and we have violation of 
$4(3 + 2 \sqrt{3})/9 \approx 2.8729$, for maximally entangled state $|\Psi_{0,2}\rangle$. For all values $\beta < 1$, we have some change in 
nonlocality of quantum states, however for 
$\beta > 9/(2 (3 + 2 \sqrt{3})) \approx 0.6961$ (This value coincides with the maximum white noise tolerance as discussed above), 
we have nonlocal asymptotic states and there is no so called sudden death of nonlocality.  

Hence, we have both situations depending on parameter $\beta$, one with sudden death of quantum nonlocality but quantum states are still 
entangled and other case where states remain nonlocal and entangled.

\section{Conclusions}
\label{Sec:conc}

We have studied the behavior of bipartite entanglement under Markovian collective dephasing. 
Using a computable entanglement monotone, we have observed the freezing dynamics of entanglement which has not been studied before for qutrit-qutrit 
quantum systems. We have analyzed the dynamics of ensemble of random pure quantum states of two qutrits and have found that all of them remain NPT and 
hence entangled for all times. The specific family of states which we have studied, also exhibit freezing entanglement phenomenon. Depending on 
parameter $\beta$, the dynamics can be different. We have also examined quantum nonlocality for these states and it turned out that again depending 
upon parameter $\beta$, the initial states either loose their nonlocality at a finite time for $\beta < 0.6961$ or remain nonlocal throughout the 
dynamics for $\beta$ larger than this value. In both of these cases, provided that $\beta > 1/4$, the states remain NPT, means entangled irrespective 
of whether they are nonlocal or not. We have found no evidence for time-invariant entanglement for qutrit-qutrit states and based on our results, we 
conjecture that under current dynamics, this feature of time-invariant entanglement may not exist for qutrit-qutrit systems. 

\section*{Acknowledgements}

The author is grateful to Dr. Daniel Alsina and Prof. Dr. Otfried G\"uhne for their helpful correspondence and submits his thanks to Prof. Dr. Gernot Alber 
for his kind hospitality at Technical University Darmstadt, where part of this work was done. 


\end{document}